# NgViz: Detecting DNS Tunnels through N-Gram Visualization and Quantitative Analysis

Kenton Born
Kansas State University
kborn@ksu.edu

Dr. David Gustafson
Kansas State University
dag@ksu.edu

## ABSTRACT
This paper introduced NgViz, a tool that examines DNS traffic and shows anomalies in n-gram frequencies. This is accomplished by comparing input files against a fingerprint of legitimate traffic. Both quantitative analysis and visual aids are provided that allow the user to make determinations about the legitimacy of the DNS traffic.

### Categories and Subject Descriptors
C.2.3 [**Computer-Communication Networks**]: Network Operations

## 1. INTRODUCTION

Recent research has seen a large push toward the visualization of network traffic. The VisSec community uses the mind's spatial reasoning capability to recognize anomalies and patterns that cannot be easily captured through algorithms and computation. However, little work has been done in using visualization to help cyber security detect DNS tunnels.

DNS tunnels allow protocols blocked by a network policy to be encapsulated inside DNS queries and responses. Iodine [8], TCP-over-DNS [12], and Dns2tcp [5] are just a few examples of open source DNS tunnels widely available.

In 1932, Zipf [15] pioneered character frequency analysis of the English language. Shannon [12] extended his work by examining the entropy of the language. Since that time, character frequency analysis has been heavily applied to cryptography and language recognition. Recently, character frequency analysis was explored for its feasibility in detecting DNS tunnels [2]. Similar to natural language, it was shown that domains and subdomains found in DNS queries and responses often follow a recognizable fingerprint, whereas a DNS tunnels have significantly higher entropy that heavily skews the character frequencies away from the typical domain fingerprint. NgViz was created as an extension of that research, automating the visualization and analysis of DNS traffic.

NgViz performs n-gram frequency analysis on DNS traffic by comparing an input file against a fingerprint of legitimate traffic. Quantitative analysis and visual aids are combined to give the user as much information as possible to make determinations about the legitimacy of the DNS traffic.

Visualizing DNS traffic is not a new concept. The dynamic behavior of DNS was demonstrated by Ren [11] while introducing the visual metaphor *Flying Term*. However, this research focused on reflection and amplication attacks along with cache poisoning. Plonka and Barford [9] used visualization to differentiate between canonical, overloaded, and unwanted DNS queries by applying context-aware clustering. Neither approach examined the visualization of DNS tunnel traffic.

Tunnel detection has seen significant research in the past. Borders [1] explored detecting HTTP tunnels by examining request regularity, inter-request delay, bandwidth usage, and transaction size. In contrast, Crotti [3,4] and Dusi [6] attempted to detect tunnels from lower in the protocol stack by looking at the inter-arrival time, order, and size of the traffic packets. Hind [7] explored detecting DNS tunnels using artificial neural networks trained off the average packet length, average number of distinct domains, average length of packets to that domain, the average number of distinct characters in the lower level domain, and the distance between lower level domains. While many of these methods show promise, it is still vital to explore other solutions that will offer a reduction in the false positives found with the methods above. A combined approach should be used for maximum effectiveness.

## 2. METHODS

Firstly, NgViz must be given a file of domains to represent the fingerprint of legitimate traffic. This file will be used for comparisons against all other traffic files. Comparison files may supplied, or they can be generated from pcap files [10] using the included NgViz pcap parser. Many networks may have a slightly altered fingerprint of legitimate traffic. Therefore, the best approach is using traffic local to the network being analyzed.

For each unique n-gram found in an input file, a new entry into a hashmap is created. This hashmap stores the number of times each n-gram has been seen in the data set. Optionally, duplicate subdomains can be removed, analyzing the traffic only when new subdomain are seen. This is beneficial to tunnel detection, and is recommended.

Firstly, each n-gram's rank is compared to the rank of the same n-gram in the fingerprint. The distance between the two ranks is the first element used when calculating a match between the two files. It is a primary component in the calculation for *rank_match* (defined below). Secondly, the frequency of the n-gram is compared to the frequency of the same ranking n-gram in the fingerprint file (which may or may not be the same characters). This measurement is a primary component of *freq_match* (defined below), and helps determine whether the two files have a similar drop in frequency from one rank to the next. The lack of a similar Zipfian distribution is a sign that there may be a tunnel in the DNS traffic. The formula for total match is shown below:

*rank_match = (#n-grams – avg(rank diff) / #n-grams)$^a$*
*freq_match = (∑(pct of fingerprint freq) / #n-grams)$^b$*
*total_match = rank_match*x + freq_match*y    {x+y = 1}*

A tunnel might be masked by a large amount of legitimate traffic from other sources and destinations. This is mitigated in several ways. Firstly, the pcap parser can split the traffic into separate files based on either IP address or domain. Tunnels typically operate over one domain, modifying the subdomain with each request. Separating the domains into different files allows each domain to be analyzed individually. Alternatively, a DNS tunnel may be deployed over several different domains, using an algorithm similar to round-robin to switch between them. Also, multiple IP addresses may be sending legitimate traffic to a domain also being used as a tunnel, effectively masking the tunnel in the noise. By segregating the IP address, it can be analyzed without the added noise seen while analyzing a particular domain.

Along with splitting traffic by IP address and domain, it is also necessary to look at it from a temporal perspective. The IP or domain may only be part of a tunnel for a brief subset of the DNS queries/responses. Therefore, the tool provides the ability to "split" on a specified interval of data points. The results are then sorted from highest match to lowest total match, making it simple to find the traffic that is furthest from the fingerprint. An example of this would be analyzing the traffic of specific IP address every 100 queries.

However, this strategy has the pitfall of being heavily skewed by a domain containing several subdomains that all have similar labels. This will cause the frequencies of the n-grams in that label to jump disproportionality high. Because of this, an option to graph the drop in frequency from one n-gram to the next was added. While this scenario will be highlighted as a poor match when viewing ranks and frequencies, it will also show several large spikes in the "change in frequency" graph that are not seen in tunnels or standard DNS traffic (see Figure 1).

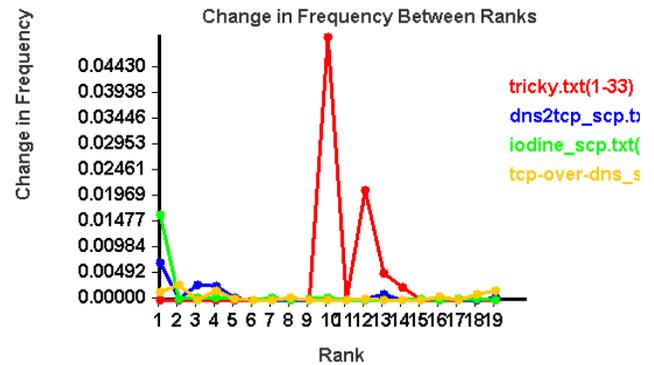

**Figure 1: Spikes resulting from many subdomains with a similar label**

## 3. RESULTS

A fingerprint file of legitimate traffic was built using Alexa's list of the 1,000,000 most popular domain names for 2009 [14]. This is postulated to provide a better representation of DNS n-gram distributions than English, which may be skewed by prepositions and pronouns that are not seen as often in subdomains. A custom website crawler was used to simulate "typical" internet traffic. The website crawler visited thousands of popular websites, capturing queries for linked sites and ads similar to what would be seen by a typical internet user.

The first experiment split the simulated traffic every 100 unique subdomains and compared the data sets to the traffic from three popular DNS tunnels: Iodine [8], TCP-over-DNS [13], and Dns2tcp [5]. The three tunnels all showed the lowest match, ranging from a 43% match to a 35% match (see blue outlines in Figures 2 and 3, below), whereas the simulated traffic matched between 81% and 60% in all cases.

The second experiment analyzed 500 unique subdomains using bigrams instead of unigrams (comparing two character groupings instead of single characters). The results were more pronounced, showing a greater than 80% match for all simulated traffic while the tunnels matched at 26% or less. This is attributed to bigrams providing significantly more data points, decreasing the chances tunneled traffic will match domain traffic appropriately.

While bigrams showed greater results, it must be noted that each additional n-gram character requires exponential increases in computation and memory. Additionally, accuracy requires more data than is necessary with smaller n-grams because of the large range of possible matches. It is hypothesized that trigrams will offer similar results with a large enough data set. However, results using unigram and bigram analysis appear to be sufficient.

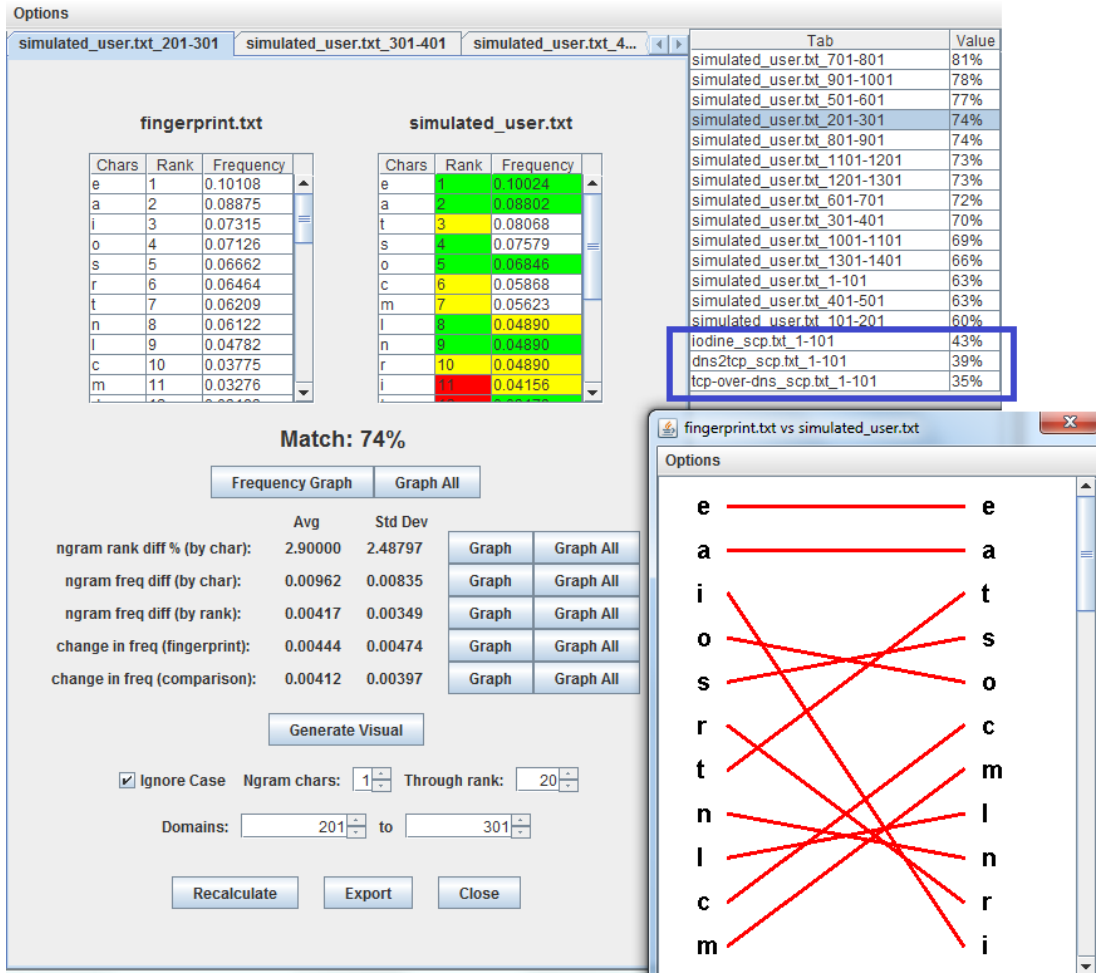

**Figure 2: NgViz unigram traffic analysis comparing simulated traffic to DNS tunnels**

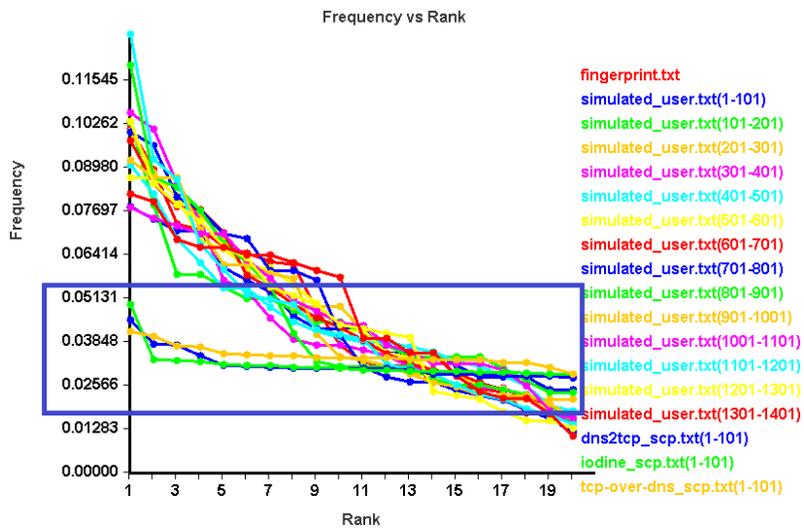

**Figure 3: NgViz frequency comparison between tunnels and legitimate traffic**

## 4. CONCLUSIONS

Combining visualization with quantitative analysis has proven to be a very effective method of detecting malicious DNS traffic. Both unigrams and bigrams were shown to provide sufficient capabilities in pinpointing non-conforming traffic.

Defeating n-gram analysis will require DNS tunnels to pad the exfiltrated data with characters that more closely match typical n-gram distributions. However, this will greatly reduce the bandwidth of the tunnel, requiring an increase in the amount of DNS traffic sent out of the malicious system.

While a match value around 50% has proved to be a good threshold for unigram analysis, it is important to note that different thresholds will be appropriate for different types of networks and different n-gram configurations. This will also vary depending on the constants *a* and *b* in the *rank_match* and *freq_match* calculations, respectively. Similarly, false positive and false negative rates are left to future research where several network data sets will be analyzed.

Future work will involve developing a real-time, automated DNS monitor that alerts administrators of anomalies in network DNS traffic. N-gram analysis will be combined with other effective methods of tunnel detection such as examining subdomain length and identifying anomalous flows out of a system on the network.

## 5. REFERENCES

[1] Borders, K. and Prakash, A. 2004. Web tap: detecting covert web traffic. In CCS'04: Proceedings of the 11[th] ACM conference on Computer and Communications Security, New York, NY, ACM Press, 110-120.

[2] Born, K. and Gustafson, D. 2010. Detecting DNS Tunnels Using Character Frequency Analysis. In Proceedings of the 9[th] Annual Security Conference, April 7-8, Las Vegas, Nevada.

[3] Crotti, M., Dusi, M., Gringoli, F., Salgarelli, L. 2007. Detecting HTTP Tunnels with Statistical Mechanisms. IEEE International Conference on Communications (ICC) '07, 6162-6168.

[4] Crotti, M., Dusi, M., Gringoli, F., Salgarelli, L. 2008. Detection of Encrypted Tunnels Across Network Boundaries. In Proceedings of the 43[rd] IEEE International Conference on Communications (ICC 2008), May 19-23, Beijing, China.

[5] Dembour, O. Dns2tcp. http://www.hsc.fr/ressources/outils/dns2tcp/index.html.en, Nov 2008.

[6] Dusi, M., Gringoli, F., Salgarelli, L. 2008. A Preliminary Look at the Privacy of SSH Tunnels. In Proceedings of the 17[th] IEEE International Conference on Computer Communications and Networks (ICCN 2008), St. Thomas, U.S. Virgin Islands.

[7] Hind, Jarod. 2009. Catching DNS Tunnels with A.I. In Proceedings of DefCon 17, July 29-Aug 2, Las Vegas, Nevada,

[8] Iodine. http://code.kryo.se/iodine/, June 2009.

[9] Plonka, D., and Barford, P. 2008. Context-aware Clustering of DNS Query Traffic. In Proceedings of the 8[th] ACM SIGCOMM Internet Measurement Conference (IMC'08), Oct 20-22, Vouliagmeni, Greece.

[10] Pcap. 2010. http://en.wikipedia.org/wiki/Pcap.

[11] Ren, P., Kristoff, J., Gooch, B. 2006. Visualizing DNS Traffic. In Proceedings of the 3[rd] International Workshop on Visualization for Computer Security, Oct 30 – Nov 3, Alexendria, VA.

[12] Shannon, C. 1951. Prediction and Entropy of Printed English. The Bell Systems Technical Journal, 30:50-64

[13] TCP-over-DNS tunnel software HOWTO. http://analogbit.com/tcp-over-dns_howto, July 2008

[14] Top Sites. http://www.alexa.com/topsites, Nov 2009.

[15] Zipf, G. 1932. Selective Studies and the Principle of Relative Frequency in Language, Cambridge, Ma.